# MODMA dataset: a Multi-modal Open Dataset for Mental-disorder Analysis

Hanshu Cai[1], Yiwen Gao[1], Shuting Sun[1], Na Li[1], Fuze Tian[1], Han Xiao[1], Jianxiu Li[1],
Zhengwu Yang[1], Xiaowei Li[1], Qinglin Zhao[1], Zhenyu Liu[1], Zhijun Yao[1], Minqiang Yang[1],
Hong Peng[1], Jing Zhu[1], Xiaowei Zhang[1], Guoping Gao[1], Fang Zheng[1], Rui Li[1],
Zhihua Guo[1], Rong Ma[1], Jing Yang[6], Lan Zhang[6], Xiping Hu[1,5],
Yumin Li*,[6], Bin Hu*[1,2,3,4]

[1] Gansu Provincial Key Laboratory of Wearable Computing, School of Information Science and Engineering, Lanzhou University, China
[2] CAS Center for Excellence in Brain Science and Intelligence Technology, Shanghai Institutes for Biological Sciences, Chinese Academy of Sciences, China
[3] Joint Research Center for Cognitive Neurosensor Technology of Lanzhou University & Institute of Semiconductors, Chinese Academy of Sciences, China
[4] Open Source Software and Real-Time System (Lanzhou University), Ministry of Education, China
[5] Shenzhen Institutes of Advanced Technology, Chinese Academy of Sciences, China
[6] Lanzhou University Second Hospital, China

*corresponding author(s): Yumin Li (Liym @lzu.edu.cn), Bin Hu (bh@lzu.edu.cn)

**Abstract.** According to the World Health Organization, the number of mental disorder patients, especially depression patients, has grown rapidly and become a leading contributor to the global burden of disease. However, the present common practice of depression diagnosis is based on interviews and clinical scales carried out by doctors, which is not only labor-consuming but also time-consuming. One important reason is due to the lack of physiological indicators for mental disorders. With the rising of tools such as data mining and artificial intelligence, using physiological data to explore new possible physiological indicators of mental disorder and creating new applications for mental disorder diagnosis has become a new research hot topic. However, good quality physiological data for mental disorder patients are hard to acquire. We present a multi-modal open dataset for mental-disorder analysis. The dataset includes EEG and audio data from clinically depressed patients and matching normal controls. All our patients were carefully diagnosed and selected by professional psychiatrists in hospitals. The EEG dataset includes not only data collected using traditional 128-electrodes mounted elastic cap, but also a novel wearable 3-electrode EEG collector for pervasive applications. The 128-electrodes EEG signals of 53 subjects were recorded as both in resting state and under stimulation; the 3-electrode EEG signals of 55 subjects were recorded in resting state; the audio data of 52 subjects were recorded during interviewing, reading, and picture description. We encourage other researchers in the field to use it for testing their methods of mental-disorder analysis.

## Background & Summary

For the past decade or so, the number of mental disorder patients, especially depression patients, has grown rapidly. According to the World Health Organization 2015 statistics, the total estimated number of global diagnosed depression patients reached 322 million [1] and increased by 18.4% between 2005 and 2015[2]. Major depressive disorder (MDD) has become a leading contributor to the global burden of disease. However, currently, the method of diagnosis depression is based on interviews, and clinical scales carried out by professionals, such as psychiatrists and psychologists. The process is not only labor-consuming but also time-consuming. The result of depression diagnosis is also not as convincing as some other illness, such as hypertension and heart disease, due to its lack of physiological indicators. Not to mention, most countries are suffering from a lack of psychiatrically and physiological doctors, especially the less developed countries. All these reasons are causing the global population to still widely undiagnosed and untreated for their mental health disorders.

With the rising of tools such as data mining and artificial intelligence, using physiological data to explore new possible physiological indicators of mental disorder and creating new applications for mental disorder diagnosis has become a new research hot topic.

Electroencephalography (EEG), as a non-invasive physiological data, provides a direct measure of postsynaptic potentials with millisecond temporal resolution. Since mental disorders, such as depression, are complex brain cognitive disfunction, EEG is naturally the common data that are favored by the researchers. Acharya et al. [3] proposed a technique that can learn automatically and adaptively from the input EEG signals to differentiate EEGs obtained from depressive and normal subjects. And it was discovered that the EEG signals from the right hemisphere are more distinctive in depression than those from the left hemisphere. Allen et al. [3] found frontal EEG asymmetry may serve as a biomarker of depression risk. It can predict future negative emotions and potentially predict treatment response. Tement et al. [3] focused on the analysis of EEG alpha frequency and suggested burnout was associated with alpha power, whereas depression was linked to individual alpha frequency. Whitton et al. [3] used high temporal resolution of EEG to compare the spectral properties of resting-state functional connectivity in individuals with major depressive disorder to healthy controls, and discovered that elevations in high-frequency default mode network and the fronto-parietal network connectivity may be a neural marker linked to a more recurrent illness course.

Audio is another non-invasive accessible physiological data, and studies have shown that mental disorders will be causing the patients' audio data to differ from healthy controls. Harati et al. [3] built their predictive model on the top of emotion-based features to help clinical management decisions during Deep Brain Stimulation treatment of major depressive disorder patients. Cummins et al. [3] studied speech-based depression classification using gender dependant features and classifiers and revealed gender differences in the effect of depression on vowel-level formant features. Williamson et al. [3] proposed an algorithm that estimates the articulatory coordination of speech from audio and uses these coordination features to learn a prediction model to track depression severity.

For researchers in the field, clean and good quality data is essential for their analysis results. However, good quality EEG and audio data are hard to be acquired, especially the clinical diagnosed patients' data. First of all, experiment subjects have to be properly diagnosed by professional doctors, not by self-rating scales. The reason is that although some well-recognized self-rating scales are good for self-evaluations, they are not as comprehensive as clinically diagnosed. Therefore, the ground truth has to be set by professional doctors. Secondly, the experiment has to be conducted and data has to be collected prior to patients taking any medication since the medication will cause brain activity change drastically. Last and most important, the experiment requires full cooperation from the subjects, who are already depressed. One of the symptoms of major depressive disorder patients is the lack of motivation to do anything. Therefore, it is very hard to ask a patient to cooperate through the whole experiment process, which could last hours.

Here we present a multi-modal open dataset for mental-disorder analysis. For now, the dataset includes data mainly from clinically depressed patients and matching normal controls. All our patients were carefully diagnosed and selected by professional psychiatrists in hospitals. At this stage, only electroencephalogram (EEG) and audio recording data are made publicly available. The EEG dataset includes not only data collected using traditional 128-electrodes mounted elastic cap, but also a novel wearable 3-electrode EEG collector for pervasive applications. The 128-electrodes EEG signals of 53 subjects were recorded as both in resting state and under stimulation; the 3-electrode EEG signals of 55 subjects were recorded in resting state; the audio data of 52 subjects were recorded during interviewing, reading, and picture description. Detail descriptions of each sub-dataset are listed accordingly in the following section. We encourage other researchers in the field to use it for testing their methods of mental-disorder analysis.

## Methods

Written informed consent was obtained from all participants prior to the experiment. Consent forms and study design were approved by the local Ethics Committee for Biomedical Research at the Lanzhou University Second Hospital in accordance to the Code of Ethics of the World Medical Association (Declaration of Helsinki). These methods are expanded versions of descriptions in our related work [10-13].

## Participants

*full brain 128-electrodes EEG experiment:*
53 participants include a total of 24 outpatients (13 males and 11 females; 16–56-year-old) diagnosed with depression, as well as 29 healthy controls (20 males and 9 females; 18–55-year-old) were recruited;

*pervasive 3-electrodes EEG experiment:*
55 participants include a total of 26 outpatients (15 males and 11 females; 16–56-year-old) diagnosed with depression, as well as 29 healthy controls (19 males and 10 females; 18–

55-year-old) were recruited;

*Audio experiment:*
52 participants include a total of 23 outpatients (16 males and 7 females; 16–56-year-old) diagnosed with depression, as well as 29 healthy controls (20 males and 9 females; 18–55-year-old) were recruited.

All participants had a normal or corrected-to-normal vision. Patients with major depressive disorder (MDD) were recruited among inpatients and outpatients from Lanzhou University Second Hospital, Gansu, China, diagnosed and recommended by at least one clinical psychiatrist. The normal controls (NC) were recruited by posters. The study was approved by the Ethics Committee of the Second Affiliated Hospital of Lanzhou University, and written informed consent was obtained from all subjects before the experiment began. All MDD patients received a structured Mini-International Neuropsychiatric Interview (MINI)[14] that met the diagnostic criteria for major depression of the Diagnostic and Statistical Manual of Mental Disorders (DSM) based on the DSM-IV[15]. The inclusion criteria for all participants were the age should between 18 and 55 years old and primary or higher education level. For MDD patients, the inclusion criteria were the diagnostic criteria of MINI met the criteria for depression, the Patient Health Questionnaire-9item (PHQ-9)[16] score of subjects was greater than or equal to 5, and no psychotropic drug treatment having been performed in the last two weeks. For MDD patients, the exclusion criteria were having mental disorders or brain organ damage, having a serious physical illness, and severe suicidal tendencies. For NC, the exclusion criteria included a personal or family history of mental disorders. The exclusion criteria for all subjects were abused or dependent alcohol or psychotropic drugs in the past year, women who were pregnant and in lactation, or taking birth control pills. Each participant received a bonus compensation of approximately USD $16 for participating in this experiment.

## Experimental material

*full brain 128-electrodes EEG experiment:*
Task 1: Resting state
No experimental material.
Task 2: Dot probe
The dataset we present is composed of facial pictures from the standardized native Chinese Facial Affective Picture System (CFAPS)[17]. The facial pictures were chosen and classified into four sets as fear, sad, happy, and neutral emotion based on their valence. Two different valences of facial pictures (one belonging to emotional sets, the other to neutral set) were selected arbitrarily. The stimuli pairs consisting of emotional and neutral facial pictures, appeared side by side on the screen. The distance between the two facial pictures was 12 cm, with a constant viewing angle of 14.25°. We obtained 60 emotional-neutral face pairs, including 20 fear-neutral, 20 sad-neutral, and 20 happy-neural faces. The number of male and female pictures of each emotion was equal. The picture size was 5.16 cm × 5.95 cm, and all non-facial features were trimmed (i.e., hair or clothing). The

MATLAB software was used to equate mean pixel luminance, contrast, and centro-spatial frequency of all face pictures, and the pictures were converted into 8-bit greyscale images.

*pervasive 3-electrodes EEG experiment:*
Only resting-state EEG was recorded; therefore, there is no experimental material used.

*Audio experiment:*
In the design of the experiment, two factors were examined: speaking style and emotional valence. Speaking style is about three speaking patterns that were involved in the study: interview, words reading, and picture description. Each of them had three kinds of emotional valences: positive, neutral and negative. The order of speech with different emotional valences was assigned randomly to counteract the sequence effect. The language of the experiment was Chinese and the whole experiment lasts about 25 minutes.
More details of the three speaking styles are described as follows:
    1) The interview had 18 questions that came from DSM-IV [15], HRSD [18], and other scales.
    2) Reading comprised two groups of words, and each group has ten common Chinese words that came from affective ontology corpus [19] and Chinese sentimental words extremum table [20].
    3) The picture description contained three facial expression pictures, which were from the Chinese Facial Affective Picture System (CFAPS) [21]. More details for experimental procedures could be found in [22].

## Experimental equipment

*full brain 128-electrodes EEG experiment:*
Continuous EEG signals were recorded using a 128-channel HydroCel Geodesic Sensor Net (Electrical Geodesics Inc., Oregon Eugene, USA) and Net Station acquisition software (version 4.5.4). The sampling frequency was 250 Hz. All the raw electrode signals were referenced to the Cz. For each participant, we first measured their head circumference and then selected the appropriate size EEG net. The impedance of each electrode was checked before recording, to ensure good contact, and was kept below 50 kΩ[18].

*pervasive 3-electrodes EEG experiment:*
Data was recorded by a 24-bit A / D converter record EEG signals at a sampling frequency of 250Hz, as shown in Figure 1.

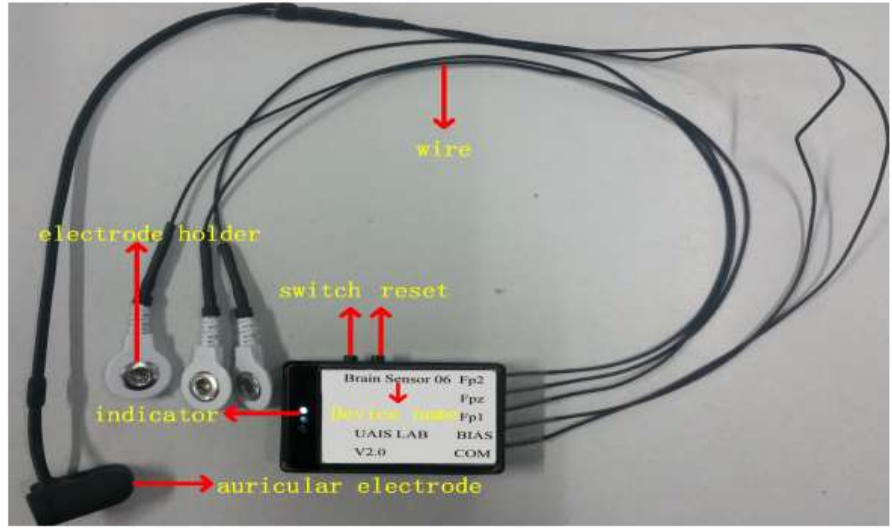

Fig 1. Three-electrode pervasive EEG collection device

*Audio experiment:*
The experiments were performed in a quiet, clean, soundproof, and no electromagnetic interference room. During the experiment, the ambient noise of the lab must be less than 60dB. The devices we used for recording are Neumann TLM102 (microphones) and RME FIREFACE UCX (audio card) with a 44.1 kHz sampling rate and 24-bit sampling depth. All recording data were saved as uncompressed W A V format.

The whole experiment lasted about 25 minutes for one participant. During recording, the subject was asked not to touch any equipment and keep the distance between mouth and microphone about 20cm. Each subject is invited to complete all three experimental tasks on a comfortable chair. Ambient noise signals were required un-der 60dB to prevent interference with the subject's audio signals.

## Experimental paradigm

*full brain 128-electrodes EEG experiment:*
The multi-channel EEG was collected in a quiet, sound-proof, well-ventilated room without strong electromagnetic interference. Participants completed the tasks sitting alone in the room, while the operators were monitoring their progress in the adjoining room. When the electrode placement is completed, and the impedance meets the requirements, data acquisition can be started. All participants were asked to complete two tasks: resting state and dot-probe tasks.

Task 1: Resting state
5 minutes of eyes-closed resting-state EEG was recorded. Participants were required to keep awake and still without any bodily movements including heads or legs, as well as any unnecessary eye movements, saccades, and blinks. After completion of Task 1, participants had a rest and then completed Task 2.

Task 2: Dot probe

Participants were seated in front of the monitor (17" monitor, 1280 ×1024 resolution, and 60 Hz refresh rate) at a distance of 60 cm. All relevant instructions were shown on the computer screen initially. Before the experiment began, the participants were instructed to complete the 10 practice trials to get familiar with the task. In the formal experiment, the participants were instructed to focus their attention on the emotional-neutral face pairs with eyes viewing freely. And they were asked to press the button on the reaction box as quickly and accurately as possible when the dot appeared. The participants must press down the button without any bodily movements including heads or legs, as well as any unnecessary eye movements, saccades, and blinks. After completing each block, they would have a rest.

The whole experimental paradigm was programmed by E-prime v2.0 (Psychology Software Tools, Inc., Pittsburgh, PA, USA). The task consisted of three blocks (Fear-Neutral, Sad-Neutral, and Happy-Neutral), and each block had 160 trials. At the beginning of each trial, a fixed white cross appeared on the central screen at 300 ms and lasted for 300 ms from the start. Then, the cross was presented on the screen centrally on the screen throughout the experiment. The emotional-neutral face stimuli pair was presented on the screen as a cue for 500 ms, the pair was arranged in a pseudo-random order. After a short interval from 100–300 ms, the dot-probe appeared randomly as a target on the left or right position of the fixed cross for 150 ms. Concurrently, the participant was asked to identify the spatial location of the 'dot' and to record their response by pressing the button '1' or '4' on the reaction box with their index fingers as quickly as possible. If the dot appeared to the left of the fixation cross, the subject should press '1'; if the dot appeared to the right of the fixation cross, the subject should press '4'. An automatic interval of 2000 ms was used to receive the response of the participant; otherwise, the participant would be directed into the subsequent trial that was followed by a black screen presented for 600 ms. The procedure continued gradually until a block was completed. Each block was also run in a cycle manner until the entire task was finished. The whole experimental task was completed in about 25 min. The trial sequence of the dot-probe task is illustrated in Fig. 2.

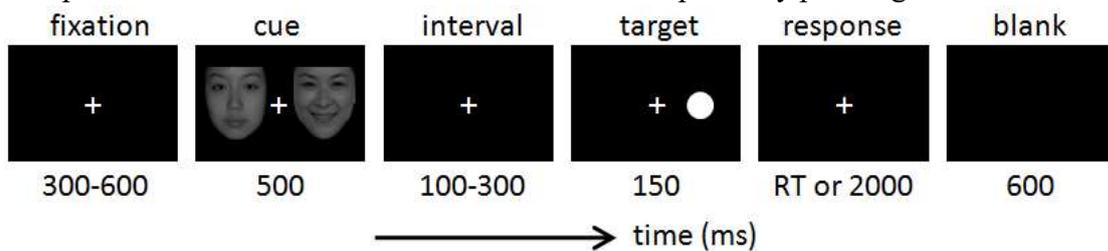

Fig. 2. The trial sequence of the dot-probe task. The cue stimuli include three kinds of emotional-neutral face pairs (Fear-Neutral, Sad-Neutral, and Happy-Neutral). The dot target was presented randomly as a target in either the left or right position of the fixed cross.

*pervasive 3-electrodes EEG experiment:*
The data recorded in a room without loud noise and strongly magnetic. Participants kept their eyes closed until they were observed their EEG signals were relatively stable, then we started a 90-second data acquisition

*Audio experiment:*

There are three fixed-order parts: interview, reading, and picture description. The text materials were showed on a computer screen, and the participants were asked to finish the experiment following the instructions.

1) Interview: This task contained 18 questions with positive, neutral and negative meanings. These topics came from DSM-IV and some depression scales which are often used in this field. For example: If you have a vacation, please describe your travel plans [22]. What is your best gift you have ever received and how did you feel [23]? Please describe one of your friends, including age, job, characters, and hobbies. How do you evaluate yourself? What would you like to do when you are unable to fall asleep? What makes you desperate?

2) Reading: This part consists of a short story named "The North Wind and the Sun", which is from the booklet "The Principles of the International Phonetic Association" [24], and often used in the acoustic analysis in international, multilingual clinical research. And three groups words with positive (e.g., outstanding, happy), neutral (e.g., center, since) and negative (e.g., depression, wail) emotion. Positive and negative words are selected from affective ontology corpus created by Hongfei Lin [25], and neutral ones are picked out from Chinese affective words extremum table [26]. All these words are often-used words in Chinese to avoid the impact of educational level and three groups words have close stroke numbers. Subjects are told to read a story, and these words in their common ways.

3) Picture description: The materials for this task include four pictures in all. Three pictures, which express positive, neutral, and negative faces, are selected from Chinese Facial Affective Picture System (CFAPS) and the last one with a "crying woman" came from Thematic Apperception Test (TAT) [23]. TAT is created by Murray in 1935, which is used in psychological counseling and psychotherapy at present. In this task, subjects are told to describe these four pictures freely.

## Data Records

### 1. Data recording and storage

*full brain 128-electrodes EEG experiment:*
Task 1: Resting state
Five minutes of eyes-closed resting data were recorded with Net Station acquisition software. The acquired raw data were saved as .mff files on the MAC PC. The data files named with "0201" prefix represent data from patients with MDD, and the data files named with "0203" prefix represent data from NC. Then .mff files were converted to .mat files using the Net Station Waveform Tools.
Task 2: Dot probe
During the experiment, the stimulus computer presented the dot-probe experiment task and recorded the reaction time (RT), accuracy and CellNumber in a .edat file. The data files named with "Dot_Detection-0201" prefix represent data from patients with MDD, and the data files named with "Dot_Detection-0203" prefix represent data from NC. The event file can be imported by E-prime. The stimulus computer also sent synchronized triggers to the Net Station acquisition software. Concurrently, the Net Station acquisition software recorded EEG data with the timestamps of triggers. The acquired raw data were saved as .mff files on the MAC PC. The data files named with "0201" prefix represent data from patients with MDD, and the data files named with "0203" prefix represent data from NC. Then .mff files were converted to .raw files using the Net Station Waveform Tools.

For the two experimental tasks, EEG signals were obtained using the HydroCel Geodesic Sensor Net (HCGSN) (Electrical Geodesics Inc., Oregon Eugene, USA). Recorded EEG signals were collected using a wired EEG cap with 128-Ag/AgCl electrodes. The contact impedance between all electrodes and the skin was kept below 50 kΩ. The EEG recordings were amplified by the Electrical Geodesics amplifiers and digitized at 250 Hz. Net Station acquisition software 4.5.4 is the ultimate tool for data acquisition.

The data are stored in folders by task, name as "128-channel_RestingState" and "128-channel_DotProbe," respectively. One file per subject.

*pervasive 3-electrodes EEG experiment:*
Considering that the prefrontal lobe has a strong correlation with emotional processes and psychiatric disorders, we collected the EEG signal by three-electrode. pervasive EEG collection device which has three electrodes located on the prefrontal lobe (Fp1, Fpz, and Fp2). The location of the three electrodes placement (Fp1, Fpz, and Fp2) and the three-electrode pervasive EEG collection device are shown in Fig. 3.

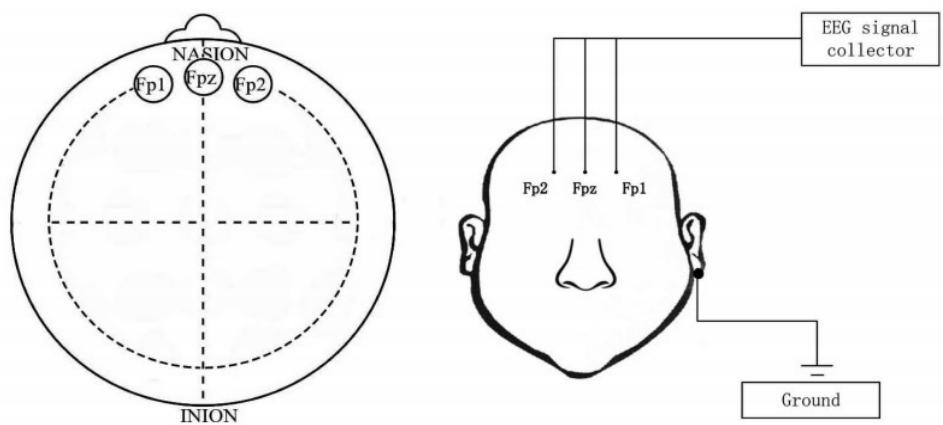

Fig. 3. Location of the three electrodes placement and the three-electrode pervasive EEG collection device

Collected data files were stored in a referential montage using an open-source TXT format. Each TXT is an M by N array, M is the number of electrodes (M = 8), and N is the number of all sample dots. What needs to be explained here is the first three electrodes correspond to the Fp1 electrode, Fpz electrode, Fp2 electrode, and the last five electrodes are alternate channel data which is the default value if not be used.

*Audio experiment:*
We collected data in a quiet, clean, soundproof, and no electromagnetic interference room. During the experiment, the ambient noise of lab must be less than 60dB. The devices we used for recording are Neumann TLM102 (microphones) and RME FIREFACE UCX

(audio card) with a 44.1 kHz sampling rate and 24-bit sampling depth. All recording data were saved as uncompressed WAV format.

All the recordings were segmented and labeled manually, and only participants' speech was kept. There were 29 recordings (interview (18), passage reading (1), word reading (6) and picture description (4)) for each subject.

## 2. 3-electrodes EEG signals

During processing, we converted the raw hex data of the first three columns to decimal data first. Then, the signal was filtered by 1Hz high-pass and 45Hz low-pass finite impulse response (FIR) filter. Next, we use an adaptive noise canceller to remove the eye-blink artifacts. Figure 4 shows the signal waveform in real-time after preprocessing.

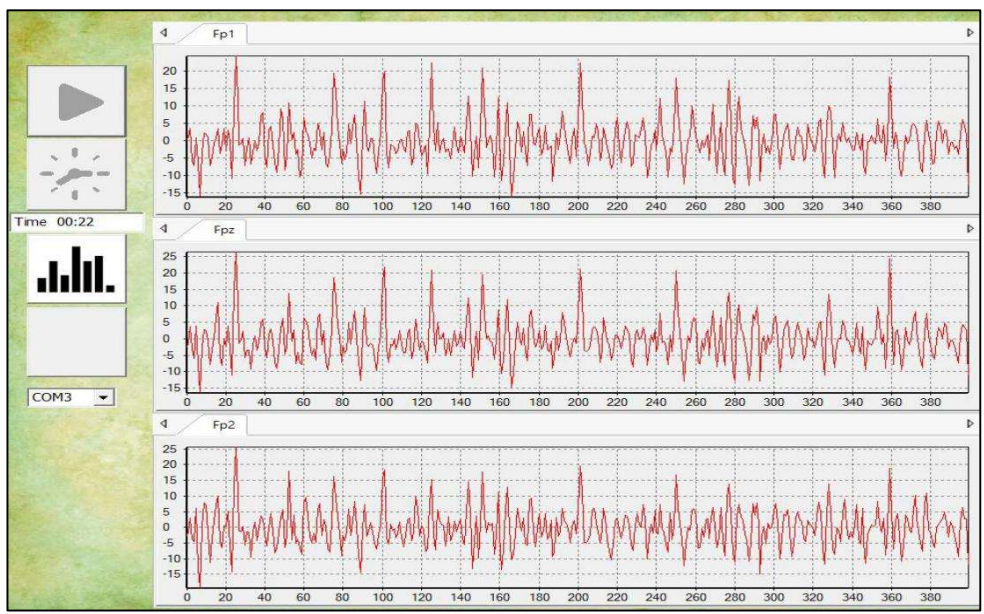

Fig. 4    The real-time graphic display system shows the EEG signals after processing in real-time.

## 3. Whole-brain EEG signals

Task 1: Resting state
The raw files were read using the EEGLAB toolbox in MATLAB. The uploaded files named with mat suffixes contain all the signals. After loading the files, the "EEG.data" variable included 129 EEG signals. The first 128 signals were from the electrode E1 to electrode E128. The last signal from Cz was the reference electrode.
Task 2: Dot probe
The raw files were read using the EEGLAB toolbox in MATLAB. The uploaded files named with raw suffixes contain all the signals. After loading the files, the "EEG.data" variable included 128 EEG signals. The 128 signals were from the electrode E1 to electrode E128. Additionally, as shown in Table 1, the types of events (see "EEG.event.type") in the dataset were classified as fixation onset (mark: hfix or ffix or sfix), cue onset (mark: hcue or fcue or scue), interval onset (mark: hisi or fisi or sisi), target onset (mark: hdot or fdot or sdot) and response onset (mark: hwrp or fwrp or swrp).

Table 1. The timestamps of triggers of different experiment blocks in the dot-probe task

| Happy-Neutral block | | | | |
|---|---|---|---|---|
| 'hfix' | 'hcue' | 'hisi' | 'hdot' | 'hwrp' |
| fixation onset | cue onset | interval onset | target onset | response onset |
| Fear-Neutral block | | | | |
| 'ffix' | 'fcue' | 'fisi' | 'fdot' | 'fwrp' |
| fixation onset | cue onset | interval onset | target onset | response onset |
| Sad-Neutral block | | | | |
| 'sfix' | 'scue' | 'sisi' | 'sdot' | 'swrp' |
| fixation onset | cue onset | interval onset | target onset | response onset |

## 4. Audio

In the experiment, 29 recordings for every single participant were stored and named as 1 to 29 in a determined sequence. The details were as follows: The positive, neutral, and negative interview recordings are named as 1-6, 7-12 and 13-18 separately. The record of the short story is named as 19. The readings of six-word groups are named as 20-21, 22-23 and 24-25 in accordance with the sequence of positive, neutral and negative emotion. 26-28 were the picture description with the same order to the reading part. The record of TAT was numbered as 29.

# Technical Validation

## 1. Behavioral validation

*full brain 128-electrodes EEG experiment:*
For both of two tasks (Resting state and Dot probe), the EEG dataset was collected from 24 subjects with MDD and 26 NC. All participants had a normal or corrected-to-normal vision. Patients with MDD were recruited among inpatients and outpatients from Lanzhou University Second Hospital, Gansu, China. The NC were recruited by posters. At the beginning of the experiment, Each participant was given approximately 10 minutes to read the experiment prevalent instructions and fill in the participant information questionnaire. Next, the subject would wear a suitable cap and began to record the EEG signal.

For the dot-probe task, RT was defined as the time period between target onset and response onset. The trials would be rejected if RT is less than 100 ms or more than 2000 ms, and all of the trials in which the participants failed to respond would be excluded from the analyses. In addition, according to whether the dot in the target and the emotional face in the cue appears on the same side or the opposite side, we can divide the data into two conditions: emotional congruent and emotional incongruent. In this study, a total of 6 conditions were included: Happy-congruent, Happy-incongruent, Sad-congruent, Sad-incongruent, Fear-congruent, and Fear-incongruent. The condition information was save in the CellNumber of .edat files. The corresponding relationship of conditions is shown in Table 2.

Table 2. The CellNumber of different emotional-congruent/ incongruent condition in dot-probe

| task | |
|---|---|
| CellNumer | Emotional-congruent/incongruent Condition |
| 1 | Happy-congruent condition |
| 2 | Happy-incongruent condition |
| 3 | Sad-congruent condition |
| 4 | Sad-incongruent condition |
| 5 | Fear-congruent condition |
| 6 | Fear-incongruent condition |

EEG signals were recorded using 128-Ag/AgCl electrodes elastic Cap (HydroCel Geodesic Sensor Net, HCGSN). The electrode skin interface was prepared by cleaning and rubbing the skin and then applying KCL-based conductive gel. The impedance of the electrodes was calibrated repeatedly until below 50 kΩ. EEG signals from the electro-cap were amplified using the Electrical Geodesics amplifiers (Electrical Geodesics Inc., Oregon Eugene, USA) and recorded at a sampling rate of 250 Hz.

*pervasive 3-electrodes EEG experiment:*
All subjects were assisted by professional psychologists to complete the Mini-Mental State Checklist(MMSE) as a preliminary judgment of depression. If the participant is at high risk of depression, a Simple Self-Test Depression Scale(PHQ-9) will be filled to judge the level of depression. All basic information is collected at the same time.

According to the information on the self-assessment questionnaire and the selection criteria, the candidate subjects were judged whether they met the experimental criteria or not. All the participants washed their hair under the leadership of the staff, then wear experimental equipment in a good experimental environment.

It should be noted that all subjects within the first two weeks of the experiment should not take any psychotropic drugs, and do not have any other mental illnesses or brain organ damage (such as epilepsy). Women with depression who are pregnant should be confirmed that: Women who are in lactation or taking birth control pills and who have been abused or depended on alcohol or psychotropic substances in the past one year should not participate in the experiment.

During the experiment, according to the international 10-20 system electrode placement standards, we choose 3 positions on the forehead to place the electrodes.

*Audio experiment:*
All participants are asked to sign informed consent, and they all have a certain level of education, which means that they can understand the questions and answer accurately. Besides, before participating in the experiment, the subjects were told to answer the question as realistically as possible, which makes these data authentically.

## 2. 3-electrodes EEG signals validation

We selected FP2 and FP1 leads of EEG signals of depression group (n=11)and the control group (n=11), each lead of the EEG data first through 0.5~30 Hz of FIR band-pass filter. Then, we calculate the relative power of δ(0.5~4 Hz), θ(4~8 Hz),α(8~14 Hz) andβ(14~30

Hz) rhythms of two leads EEG FP2, FP1. Using SPSS 19.0 software packages to do paired sample T-test of the left and right brain αrhythm of two groups, respectively. Statistical analysis shows that a significant difference is found in the relative power of the left and right brain signals αrhythm of the depression group of the αrhythm, with statistical significance. While the difference of relative power of the control group αrhythm does not reach a significant level, no statistical significance.

For raw public data, we have all pre-processed and identified the data quality through experiment experience.

## 3. Whole-brain EEG signals validation

EEG data of both tasks (Resting state and Dot probe) were raw data, which were saved in the MODMA dataset.

## 4. Audio validation

The experiments were performed in a quiet, clean, soundproof, and no electromagnetic interference room. Each recording was gathered following the basic process: one-minute rest, finishing the task showed on the computer screen and another break for one minute. During the experiment, if the noise is more than 60db, it will not be recorded. Besides, our good recording equipment makes the signal 'quality' good.

The 'quality' of the sound recordings is an informal and rough indication of the ratio of signal power to noise power (SNR). The SNR ranges from 20 to 30 and can be calculated normally.

# Usage Notes

The raw experimental dataset can be downloaded from the publicly accessible repository free of charge at http://modma.lzu.edu.cn/data/index/. All the users interested in this dataset will need to sign an End User License Agreement (EULA) before their downloads. The raw datasets are packaged under "EEG" and "Audio" categories respectively. Within each package, there is an excel file containing demographic data and psychological assessment scores of all the subjects of that dataset. Each experiment subject is given with an identity number and this subject id is unique across all the packages. In the future, more data will be added regularly, which will cover not only more mental disorders, such as schizophrenia, anxiety, and mania but also more data types, such as eye-movement tracking, facial expression recording, and MRIs. We encourage other researchers in the field to use it for testing their methods of mental-disorder analysis.

# Acknowledgments

This work was supported in part by the National Key Research and Development Program of China (Grant No. 2019YFA0706200), in part by the National Natural Science Foundation of China (Grant No.61632014, No.61627808, No.61210010), in part by the

National Basic Research Program of China (973 Program, Grant No.2014CB744600), and in part by the Program of Beijing Municipal Science & Technology Commission (Grant No.Z171100000117005).